\documentclass[sigconf,nonacm]{acmart}
\usepackage[ruled,vlined]{algorithm2e}
\usepackage{algorithmic}
\usepackage{multirow}
\usepackage{bbding} 
\usepackage{soul}

\SetKwInput{KwInit}{init}
\SetKwInput{KwSend}{send}
\SetKwInput{KwRecieve}{recieve}

\DeclareMathOperator*{\argmin}{arg\,min}
\AtBeginDocument{%
  }

\setcopyright{acmlicensed}
\copyrightyear{2024}
\acmYear{2024}
\setcopyright{acmlicensed}\acmConference[RecSys '24]{18th ACM Conference on Recommender Systems}{October 14--18, 2024}{Bari, Italy}
\acmBooktitle{18th ACM Conference on Recommender Systems (RecSys '24), October 14--18, 2024, Bari, Italy}
\acmDOI{10.1145/3640457.3688143}
\acmISBN{979-8-4007-0505-2/24/10}

\begin{document}

\title{Cross-Domain Latent Factors Sharing via Implicit Matrix Factorization}

\author{Abdulaziz Samra}
\email{abdulaziz.samra@skoltech.ru}
\orcid{0009-0004-9793-7832}
\affiliation{%
  \institution{Skolkovo Institute of Science and Technology}
  \city{Moscow}
  \country{Russia}
}
\author{Evgeney Frolov}
\email{frolov@airi.net}
\orcid{0000-0003-3679-5311}
\affiliation{%
  \institution{AIRI, Skolkovo Institute of Science and Technology}
  \city{Moscow}
  \country{Russia}
}
\author{Alexey Vasilev}
\email{alexxl.vasilev@yandex.ru}
\affiliation{%
  \institution{Sber AI Lab}
  \city{Moscow}
  \country{Russia}
}
\author{Alexander Grigorievskiy}
\email{alex.grigorievskiy@gmail.com}
\affiliation{%
  \institution{Comparables.ai}
  \city{Helsinki}
  \country{Finland}
}
\author{Anton Vakhrushev}
\email{btbpanda@gmail.com}
\affiliation{%
  \institution{Sber AI Lab}
  \city{Moscow}
  \country{Russia}
}

\begin{abstract}

Data sparsity has been one of the long-standing problems for recommender systems. One of the solutions to mitigate this issue is to exploit knowledge available in other source domains. However, many cross-domain recommender systems introduce a complex architecture that makes them less scalable in practice. On the other hand, matrix factorization methods are still considered to be strong baselines for single-domain recommendations. In this paper, we introduce the CDIMF, a model that extends the standard implicit matrix factorization with ALS to cross-domain scenarios. We apply the Alternating Direction Method of Multipliers to learn shared latent factors for overlapped users while factorizing the interaction matrix. In a dual-domain setting, experiments on industrial datasets demonstrate a competing performance of CDIMF for both cold-start and warm-start. The proposed model can outperform most other recent cross-domain and single-domain models. We also provide the code to reproduce experiments on GitHub. \footnote{\url{https://github.com/aa-samra/cd-imf}}
\end{abstract}



\maketitle

\section{Introduction}
\label{sec:intro}
In recent decades, recommender systems (RSs) have become increasingly popular in many online markets and social media platforms. They generate personalized suggestions for new content for users, providing them with a more engaging experience. Among recommendation algorithms, both Matrix Factorization (MF) and Neural Network approaches remain applicable in the industry. Recent studies have shown that MF algorithms are fairly strong baselines, outperforming many neural network algorithms \cite{Dacrema}. The use of MF algorithms is determined by their ease of use and speed of inference. However, such algorithms are not free of drawbacks. Data sparsity \cite{bin} and the cold start \cite{zhu2021transfer} are the common sources of problems that easily impede learning proper user and item representations and severely influence recommendation accuracy.

Large tech companies such as Amazon, Google, Facebook, and many others are actively developing their ecosystems which consist of groups of interconnected companies \cite{ecosyst}. These companies complement each other trying to cover users' interests in all item niches. To compete in the market, ecosystems include new organizations covering niches that have not yet been occupied. This renders the cold-start issue, i.e., the complete lack of user-item interactions, or sparsity issue due to a limited number of users with a narrow interactions history. These issues combined form one of the most challenging problems facing RSs in different companies, related to insufficient transactional and feedback data required for accurate inference of users’ similarities \cite{Idrissi2020}. 

Cross-domain recommendation (CDR) was proposed some years ago \cite{Zhu2021} to retain and expand the user base. The recommendation systems organizations can use data from other domains, such as related companies from the same ecosystem with a more extended history. Cross-domain recommender systems (CDRS) aim to increase the quality of recommendations by exploiting knowledge from multiple source domains \cite{Cantador2015}. This type of recommender system can help either when one domain is rich in information and the other has more sparse data (single target CDR) or when they are equally sparse (dual target CDR) \cite{Zhu2021}. In this setting, it is assumed that there is a shared set of users between related domains. However, some privacy restrictions exist for sharing user data across multiple domains in such scenarios.

Privacy concerns set a new challenge for cross-domain recommender systems. Cross-domain information transfer must occur without explicitly exposing users' history to the other domains. The shared information must be safe to transmit over insecure channels so that user data can not be extracted.
We also considered this requirement in our method, and propose to exchange a mixture of two different matrices between domains, making restoring the original data difficult.

On the other hand, a significant drawback of modern CDR methods is the requirement for resources and the complexity of implementation. Thus, we propose CDIMF; \emph{a cross-domain variant of the alternating least-squares (ALS)} \cite{Hu2008} which is well-known for its simple implementation and fast convergence.

Our main contributions to this work can be summarized as follows:
\begin{itemize}
    \item We proposed the Cross-Domain  Implicit Matrix Factorization (CDIMF) model for cross-domain recommendations based on the ALS model.
    \item The proposed algorithm outperformed state-of-the-art approaches on different pairs of datasets.
    \item The new approach can be used for the warm-start scenario, as well as for the cold-start scenario.
\end{itemize}

The structure of the paper is organized as follows. In Section \ref{sec:statement}, we articulate the problem statement we aim to address. Section \ref{sec:literature} contextualizes our work within the existing body of research on CDR. Our proposed model is introduced in Section \ref{sec:admm} while the experimental framework, including methodology and setup, is elucidated in Section \ref{sec:experiments}. Section \ref{sec:results} presents the empirical results, including comparisons with existing models and an analysis of our model's performance.  Finally, Section \ref{sec:conclusion} provides a comprehensive discussion of our findings, conclusions drawn, and prospects for future research.

\section{Problem Statement}
\label{sec:statement}
One of the famous models to predict user feedback is to learn latent factors for users and items. Let's consider the basic matrix factorization problem:
\begin{equation}
    \min_{\{X\}, \{Y\}}  F(X,Y),
\end{equation}
where:
\begin{equation}
F(X,Y)=\|C \circ  (P-X Y^T)\|_F^2 + \lambda \left(\|\Lambda_{X} X\|_F^2+ \|\Lambda_{Y} Y\|_F^2\right). \label{eq:mf_single}
\end{equation}
In this notation, $P\in \mathbb{R}^{|\mathcal{U}|\times|\mathcal{I}|}$ is the user-item interaction matrix, $X\in \mathbb{R}^{|\mathcal{U}|\times d}$ and $Y\in \mathbb{R}^{|\mathcal{I}|\times d}$ are the latent factors of size $d$ for users from the set $\mathcal{U}$ and items from the set $\mathcal{I}$, respectively. The weighting matrix $C$ is usually used to add different levels of importance to the errors $(P-X Y^T)$ in the loss function. We add the norms of the latent factors to the objective function for regularization. The hyperparameter $\lambda$ controls the regularization effect globally, and $\Lambda_X, \Lambda_Y$ are diagonal matrices used to regularize each user/item embedding individually. More details about our selection for values of $\Lambda_X, \Lambda_Y$, and $C$ are in Section \ref{subsec:als_adapt}.

We aim to extend the formula \eqref{eq:mf_single} to the case of $N$ domains, the minimization problem would be:
\begin{align}
    \min_{\{X_i\}, \{Y_i\}} \sum_{i}^{N}F_i(X_i,Y_i),
    \label{eq:mf_no_sharing}
\end{align}
where the subscript $i$ refers to the corresponding matrices on each domain. In cross-domain scenarios, having a set of overlapped users between domains can unveil hidden correlations that can help to learn a better recommendation model. For this purpose, we introduce equality constraints for all user factors across domains
\begin{align}
    \min_{\{X_i\}, \{Y_i\}} & \sum_{i}^{N}F_i(X_i,Y_i), \label{eq:mf_with_sharing} \\
    \text{s.t.} &\hspace{1em} X_1=X_2=\dots=X_N. \nonumber
\end{align}
The parameters of this objective function $\{X_i\}, \{Y_i\}$ are distributed over the $N$ domains, and that function can be jointly minimized. We hypothesize that the equality constraints would help with information sharing between domains by making the factorization process on each domain aware of the pattern in the others.

\section{Related Work}
\label{sec:literature}
Implicit Matrix factorization (IMF) model with alternating least-squares (ALS) algorithm~\cite{Hu2008} has been a widely popular recommendation model in academic research~\cite{chen2_2020}, ~\cite{He2016}. It has also been a workhorse model in many industrial applications due to its simplicity, scalability, and strong performance~\cite{implicit}. Despite the invention of more powerful deep-learning models, IMF has proven to be a strong baseline model for collaborative filtering recommender systems, provided hyper-parameters are properly tuned~\cite{Rendle2021}. However, it implies that all the user data is located on one server (or cluster) and that data is coming from a single domain. Our algorithm addresses these ALS issues. It allows each data provider to keep the user data private and exchange only perturbed user embedding representations. Moreover, by putting different weights on different data domains, the model extends to the CDR setup.


The Alternating Direction Method of Multipliers (ADMM) has been successfully applied in recommender systems design before. For instance, ~\cite{steck2020} considers ADMM as a convenient optimization algorithm for the SLIM~\cite{ning2011} model which is a sparse linear autoencoder for collaborative filtering. Research in ~\cite{yu2014}, ~\cite{hue2019}, ~\cite{zhang1_2022} apply ADMM to optimize IMF objective function in a high-performance computing environment. Hence, their goal is to address the large-scale problem and to shorten the model training time. Furthermore, ~\cite{hue2019}, ~\cite{zhang1_2022} utilize the differential privacy technique to prevent uncovering user identities. However, the model is still trained on a centralized cluster, which implies access to the entire dataset. Overall, previous usage of ADMM is unrelated to cross-domain setup and restrictions on sharing user interaction data.

A recent overview of CDRS is provided in~\cite{zhang2_2022}. The proposed taxonomy of methods consists of two levels. The first level comprises various degrees of user data overlap between domains. The second level considers different recommendation tasks that can be posed on CDRS. On the first taxonomy level, our formulation is applicable when there is no item overlap and full user overlap between domains. Following the second level, our formulation can be considered as both intra-domain and inter-domain formulation. Also, it is a multi-target system since recommendations can be made for all involved domains.

Before the deep learning era, CDRS was mostly based on linear matrix factorization models. For example, work~\cite{weike2010} proposes an IMF model to learn user and item embeddings on the source domain. Target domain embeddings are found to be close to the source embeddings in a Frobenious norm sense. In the article~\cite{man2017} a linear (also nonlinear) mapping function between source and target domain embeddings is proposed. In contrast, our algorithm does not separate source and target domains thereby being a multi-target algorithm. Collective matrix factorization (CMF) based algorithms~\cite{singh2008},~\cite{zhao2018},~\cite{zhang2_2022} has been also applied to the CDR problem. In contrast to our approach, Stochastic Gradient Descent is typically used for solving the CMF optimization problem. Our algorithm is an extension of the ALS algorithm which makes it familiar to the broad academia and industry communities. It is capable of reproducing the exact single-domain ALS solution when hyperparameters are set properly.

A vast amount of deep learning models have been proposed for CDR~\cite{zhang2_2022}. 
Motivated by the recent work of Cao et al. \cite{Cao2023}, who proposed a universal CDR model, we aim to develop a model that can be applied to multiple CDR scenarios. Unlike UniCDR, which adopts a neural contrastive learning approach, we will extend a standard MF model to the cross-domain setting to inherit the advantages of simplicity and computational efficiency.

\section{Proposed Model}
\label{sec:admm}
To achieve knowledge transfer across domains, we aim to make the factors of the shared users equal over all domains at the end of training. A famous method to train a distributed model with shared parameters is applying the Alternating Direction Method of Multipliers (ADMM). It is an algorithm that is used to solve convex optimization problems. It works by breaking the problem into smaller subproblems that can be solved on separate sides \cite{Boyd2010}. ADMM sharing optimization problem is:

\begin{align}
    \min_{x_i,z} & \sum_{i=1}^{N}f_i(x_i) + g(z),\label{eq:basic_admm} \\
    \text{s.t.} &\hspace{1em} x_i-z=0. \nonumber
\end{align}
    
This problem is called \emph{Global Variable Consensus with Regularization} \cite{Boyd2010}. It is about minimizing the functions $\{f_i(x)\}$ located on different sides while trying to keep the values of $\{x_i\}$ as close as possible to the shared variable $z$. Iterations for solving this problem are defined as follows:
\begin{align}
    x_i^{k+1} &= \argmin_x f_i(x) + \frac{\rho}{2} \|x-z^k +u_i^k\|_2^2, \\ 
     z^{k+1} &= \argmin_z g(z) + \frac{N \rho}{2} \|z-\left<x\right>^{k+1}-\left<u\right>^k\|_2^2, & \\
    u_i^{k+1} &= u_i^k + (x_i^{k+1} -z^{k+1} ).
\label{eq:basic_iter} 
\end{align}
In these equations, $k$ is the iteration number, $u_i$ is the dual variable related to the equality constraint $x_i-z=0$, and $\rho$ is the penalty parameter that controls how strong the local variable $x_i$ is attracted to the global $z$ in every iteration.

Our proposed approach is to get common user factors over domains while factorizing interaction matrices. Then, we rewrite the optimization problem \eqref{eq:mf_sharing} as:

\begin{align}
    \label{eq:admm_opt}
    \min_{\{X_i\}, \{Y_i\}, Z} &  \sum_{i}^N F_i(X_i,Y_i)  + g(Z), \\
    \text{s.t.} & \hspace{1em} X_i-Z=0. \nonumber
\end{align}

\subsection{Update Iterations for CDIMF}
Let us write the Augmented Lagrangian for the optimization problem \eqref{eq:admm_opt}
\begin{equation}
    \mathcal{L}\left(\{X_i\},\{Y_i\},Z,\{U_i\}\right) = F_i(X_i,Y_i) + \frac{\rho}{2} \|X_i-Z +U_i\|_F^2  + g(Z).
\label{eq:mf_sharing}
\end{equation}
As mentioned before, $\rho$ is a penalty parameter. Experiments (Section \ref{subsec:rho_effect}) will demonstrate that this parameter controls the speed of information sharing. Similarly, variables $\{U_i\}$ are the dual variables for the equality constraints $X_i-Z=0$. The iterations for solving the federated matrix factorization problem using ADMM, are:
\begin{align}
    X_i^{k+1}, Y_i^{k+1} &= \argmin_{X_i,Y_i} F_i(X_i,Y_i) +\frac{\rho}{2} \|X_i-Z^k +U_i^k\|_F^2,   \label{eq:admm_update_local}\\
     Z^{k+1} &= \text{prox}_{g,\frac{1}{\rho N}}\left(\left<X\right>^{k+1}+\left<U\right>^k\right), \label{eq:admm_update_global}\\
    U_i^{k+1} &= U_i^k + (X_i^{k+1} -Z^{k+1}).
\end{align}
Here, $N$ is the total number of domains, and $\text{prox}_{g,\mu}$ is the proximal operator of the regularization function $g(Z)$. We can also consider $g(Z)$ as a regularization term for the shared variable $Z$, and several regularization patterns can be used in the aggregation step \eqref{eq:admm_update_global}. Finally, these brackets $\left<.\right>$ denotes the mean of the corresponding variable over all the domains. 

\subsection{Proximal Operator Options}
The proximal operator \cite{parikh2014proximal} for a function $g(X)$ is given by the formula:
\begin{equation}
    \text{prox}_{g, \mu}(Z) = \argmin_X g(X) + \frac{1}{2\mu} \|X-Z\|_F^2.
    \label{eq:proximal_basic}
\end{equation}
This operator solves a double-target optimization problem. It minimizes the function $g(X)$ while trying to keep $X$ as close as possible to some value $Z$, and the parameter $\mu$ controls this trade-off.
In this paper, we consider two simple cases of proximal operators to use in the step \eqref{eq:admm_update_global} of the ADMM iterations, and here we list these options: 
\subsubsection{Identity Operator}
For the case of $g(X)=0$, the proximal operator will be: 
\begin{equation}
    \text{prox}_{0, \mu}(Z) = Z.
\end{equation}
This is a straightforward option that was used in many previous works \cite{Yan2019, Wang2022, Quoc2021}. 
\subsubsection{L2 Regularization}
For $g(X)=\frac{\lambda}{2}\|X\|_F^2$, which is the squared Frobenius norm, the proximal operator is: 
\begin{equation}
    \text{prox}_{\|.\|_F^2, \mu}(Z) = \argmin_X \frac{\lambda}{2}\|X\|_F^2 + \frac{1}{2\mu} \|X-Z\|_F^2. 
\end{equation}
Solving this unconstrained problem is easy,  $\frac{\partial f}{\partial X} = \lambda X + \frac{1}{\mu} (X-Z) = 0$. Then, 
\begin{equation}
    \text{prox}_{\|.\|_F^2, \mu}(Z) = \frac{\mu}{1+\lambda\mu}Z. 
\end{equation}
However, more options can be used in this step, like projection to a specific set of matrices, or volume maximization. These options are left for future research.

\subsection{Details of the used ALS solver} 
\label{subsec:als_adapt}
We adapt the solver proposed by Rendle et al. \cite{Rendle2021} to solve the optimization problem in step \textbf{4.} of the Algorithm (\ref{alg:admmNsides}). We chose that notation of ALS since it helps to guide the hyperparameter search iteratively rather than jointly. Rendle et al. used the following objective function for the ALS problem:
\begin{equation}
    L(X,Y) = L_S(X,Y) + L_I(X,Y) + R(X,Y), \label{eq:als_total_loss}\\
\end{equation}
\begin{align}
    L_S(X,Y) =& \sum_{(u,i)\in \mathcal{S}} (x_u^T y_i-1)^2, \\
    L_I(X,Y) =& \alpha \sum_{u\in U}\sum_{i\in I} (x_u^T y_i)^2,\\
    R(X,Y)=&\lambda\left(\sum_{u\in U}(|\mathcal{I}_u|+\alpha |\mathcal{I}|)^v\|x_u\|^2+\sum_{i\in I}(|\mathcal{U}_i|+\alpha |\mathcal{U}|)^v\|y_i\|^2\right).
\end{align}

Where $x_u, y_i \in \mathbb{R}^d$ are the latent factors for the user $u$ and the item $i$, respectively. The loss objective in \eqref{eq:als_total_loss} consists of 3 terms; the first term $L_S$ is the squared error between the predicted score and the observed interaction for all the pairs (user/item) in the dataset $\mathcal{S}$. The next term $L_I$ is computed for all pairs in $\mathcal{U}\times\mathcal{I}$ and represents the difference of the predicted score from 0. This loss helps the model not to learn a naive predictor $x_u y_i = 1,\forall (u, i)\in\mathcal{U}\times\mathcal{I}$. The hyperparameter $\alpha$ controls the contribution of $L_I$ in the total loss function. The last term is the regularization term $R$, which is controlled by the global regularization parameter $\lambda$. 

The scheme of individual regularization weights depends on the number of observed interactions for a specific user $|\mathcal{U}_i|$  or item $|\mathcal{I}_u|$  and on the size of the users set $|\mathcal{U}|$ and items set $|\mathcal{I}|$. The scheme is controlled by the frequency-scaled regularization parameter $v$. This parameter takes values from the range $[0,1]$. For $v=0$, all user and item factors will be regularized equally. The larger $v$ is, the more variance there will be in the regularization effect between popular and unpopular items/users. 


The following substitutions are required to adapt this model to our notation \eqref{eq:mf_single}:

\begin{align}
    C =& \begin{cases}
            1+\alpha, & \text{if } (u,i)\in S \\
            \alpha, & \text{if } (u,i)\notin S
        \end{cases}, \\
    \Lambda_X =& \text{diag}\left(\{|\mathcal{I}_u|+\alpha |\mathcal{I}|)^v ; u \in U\}\right), \\ 
    \Lambda_Y =& \text{diag}\left(\{|\mathcal{U}_i|+\alpha |\mathcal{U}|)^v ; i \in I\}\right).
\end{align}

\subsection{Derivation of equations for the iterations of the ALS solver}
\label{subsec:als_derivation}
We will derive the solution of a general form that can be adapted to solve multiple steps of the CDIMF algorithm with proper substitutions. The generic formula is:
\begin{equation}
    A = \argmin_{A} \frac{1}{2}\|C \circ (P-AB^T)\|_F^2 + \frac{\lambda}{2} \|R A\|_F^2 + \frac{\rho}{2} \|A-H\|_2^2.
\end{equation}
This problem is convex with respect to $A$, so the optimal $A$ is the solution of $\partial f/\partial A = 0$, then:
\begin{equation}
    \frac{\partial f}{\partial A} = - \left[C^{\circ 2} \circ (P-AB^T) \right] B + \lambda R^2 A + \rho (A -H) = 0,
\end{equation}
where $C^{\circ 2}$ is the element-wise square. We can solve the problem for a single column in $A$, let us denote $a, h$ for each \emph{row} of matrices $A, H$. We also use $p, c$ for a \emph{row} of $P, C$ respectively, and $r$ is a scalar since $R$ is a diagonal matrix.  
$$- \left[c^{\circ 2} \circ (p - a_{row} B^T) \right] B + \lambda r^2 a_{row} + \rho (a_{row} -h_{row}) = 0,$$
$$- \left[(p_{row} - a_{row} B^T) \circ c_{row}^{\circ 2} \right] B + \lambda r^2 a_{row} + \rho (a_{row} -h_{row}) = 0,$$
$$\left[(a_{row}B^T) \circ c_{row}^{\circ 2} \right] B + \lambda r^2 a_{row} + \rho a_{row}  = (p_{row} \circ c_{row}^{\circ 2}) B + \rho h_{row},$$
$$ a_{row} B^T \text{diag}(c_{row}^{\circ 2}) B + \lambda r^2 a_{row} + \rho a_{row}  = (p_{row} \circ c_{row}^{\circ 2}) B+ \rho h_{row}.$$
Transpose: 
\begin{equation}
 \left(B^T \text{diag}(c^{\circ 2}) B + (\rho + \lambda r^2) I_d \right) a  = B^T (p \circ c^{\circ 2}) + \rho h. \label{eq:als_lin_sys}
\end{equation}
Solving this linear system updates a single user/item factor. Note that the problem \eqref{eq:als_lin_sys} for each factor is independent of the other factors, and this feature allows great possibility for parallelism.
With a very similar derivation, we can get a similar formula to update $B$. So, here we list substitutions for solving different problems in the steps of the algorithm (\ref{alg:admmNsides}):
\begin{itemize}
    \item For shared users: $A=X, B=Y, H=Z-U$
    \item For non-shared users: $A=X, B=Y, H=0$
    \item For items: $A=Y, B=X, H=0$ and $p,c$ will represent the corresponding \emph{columns} of the matrices $P, C$
\end{itemize}
We remind readers that $X$ are local users' factors, $Y$ are items' factors, $Z$ are global users' factors, and $U$ are dual variables related to the equality constraints.

\subsection{Algorithm and Complexity}
The pseudocode of the algorithm running on each side (domain) is presented in Algorithm (\ref{alg:admmNsides}). In this algorithm, the matrices of user and item factors are initialized with random values from a normal distribution of standard deviation $\sigma$. This hyperparameter is recommended to be normalized by $\sqrt{d}$, where $d$ is the embedding size of the factors. It helps to control the values of the dot products of factors at the beginning of training \cite{Rendle2021}.

The complexity of one epoch for the step of updating local variables is the same as for standard ALS for a sparse matrix. It is $\mathcal{O}(d |\mathcal{S}| + d^2 |\mathcal{U}|+d^2 |\mathcal{I}|)$, where $|\mathcal{S}|,|\mathcal{U}|, |\mathcal{I}|$ are the size of observed training set, the number of users, and the number of items, respectively. For the proposed options for the proximal operator, the update of the global variable $Z$ is just the average of $X_i+U_i$ multiplied by a scalar. Thus, its complexity is $\mathcal{O}(N d |\mathcal{U}|)$, where $N$ is the number of domains. 

As illustrated in Section \ref{subsec:als_derivation}, the task of updating individual user/item factors is \textit{embarrassingly parallel}. In other words, each column of factors' matrices $X_i, Y_i$  can be computed in parallel without dependency on the others. Exploiting this feature in implementation in addition to the sparse nature of the interaction matrices can achieve significant speedup in practice \cite{Samra2024speedup}. 

\SetAlgoVlined
\begin{algorithm}[htb]
\setlength{\abovedisplayskip}{0pt}
\setlength{\belowdisplayskip}{0pt}
\caption{CDIMF psuedocode for each domain $i$}
\label{alg:admmNsides}
\KwInit{$ X_i, Y_i \sim \mathcal{N}(0,\sigma/\sqrt{d})$}
\KwInit{$U_i = 0, Z = 0$}
\For {$k \gets 1$ \KwTo $K$}{%
    \begin{flalign*}
        X_i^{k+1} \gets & \argmin_{X_i} \|C_i \circ (P_i-X_i Y_i^{kT})\|_F^2 + \|\Lambda_{Xi} X_i\|_F^2  \\
        & \hspace{5em} + \frac{\rho}{2}\cdot \|X_i-Z^k +U_i^k\|_2^2  \\
        Y_i^{k+1} \gets & \argmin_{Y_i} \|C_i \circ (P_i-X^{k+1}_i Y_i^T)\|_F^2 + \|\Lambda_{Yi} Y_i\|_F^2 \hspace{2em}
    \end{flalign*}
    
    \KwSend{$X_i^{k+1}+U_i^k$}
    \KwRecieve{$X_j^{k+1}+U_j^k \textbf{ for } j\in[1,2,\dots N]\backslash\{i\}$}
    \begin{flalign*}
        \left<X\right>^{k+1} \gets & \frac{1}{N} \sum_{i=1}^N X_i^{k+1}\\
        \left<U\right>^k \gets & \frac{1}{N} \sum_{i=1}^N U_i^k\\
        Z^{k+1} \gets & \text{prox}_{g,\frac{1}{\rho N}}\left(\left<X\right>^{k+1}+\left<U\right>^k\right)\\
        U_i^{k+1} \gets & U_i^k + (X_i^{k+1} -Z^{k+1}) \hspace{12em}
    \end{flalign*}
    
        
        
        
\KwRet{$X_i, Y_i$}
}
\end{algorithm}
\setlength{\abovedisplayskip}{5pt}
\setlength{\belowdisplayskip}{5pt}

\section{Experimental Setup}
\label{sec:experiments}
In this section, we illustrate the methodology we followed in our experiments. First, we provide details about the datasets used and the preprocessing procedure. Then, we describe the evaluation method and metrics, and finally, we list the baselines we used to compare our model with. The code to reproduce experiments is provided in the corresponding GitHub repository \footnote{\url{https://github.com/aa-samra/cd-imf}}.

\subsection{Datasets}
\label{sec:datasets}
Following many previous CDR works, we conduct experiments under the public Amazon dataset\footnote{\url{https://jmcauley.ucsd.edu/data/amazon/index_2014.html}} for a fair comparison. We used preprocessed versions of these datasets provided by \cite{Cao2023}, and these datasets were also used by \cite{liu2020cross, cao2022disencdr, cao2022cross}. However, we provided a preprocessing pipeline for further experiments. 

\begin{figure*}
    \centering
    \includegraphics[width=\textwidth]{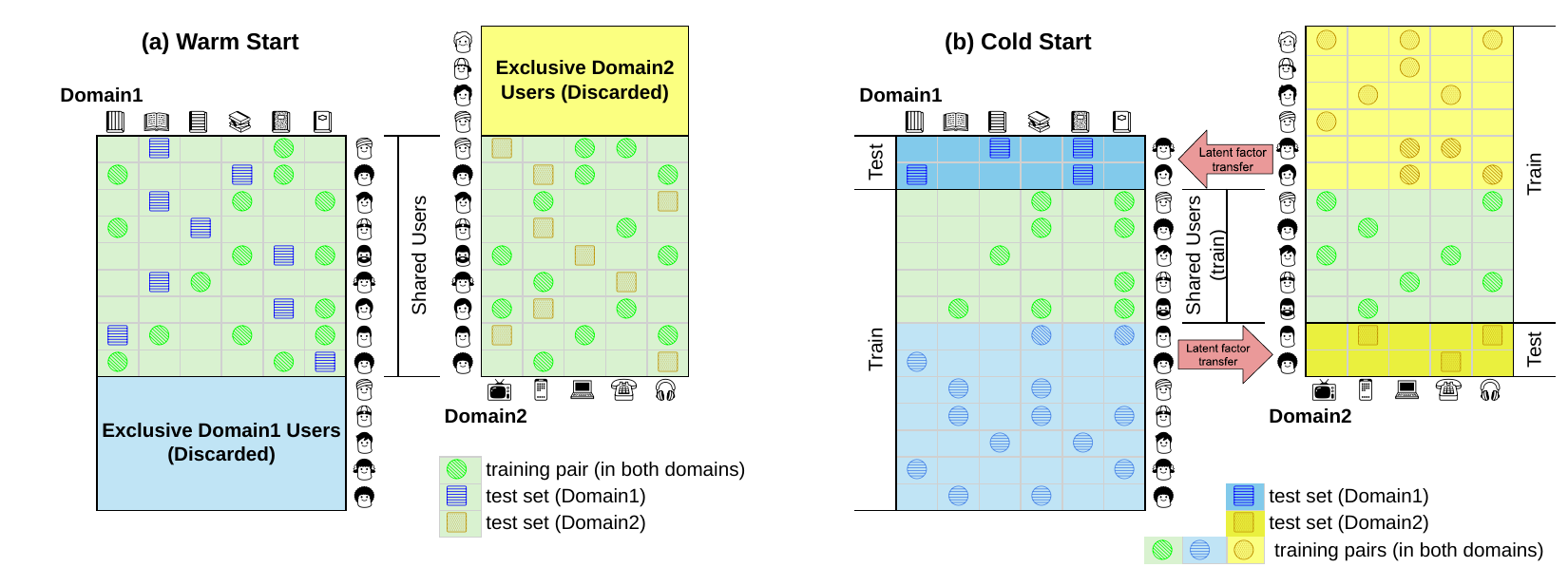}
    \caption{Train/test data splits for (a) intra-domain recommendation scheme (warm start) and (b) inter-domain recommendation scheme (cold start); circles refer to training interactions, and squares refer to test interactions. Arrows refer to latent factors transfer for testing}
    \Description[Splitting of datasets]{Splitting of datasets}
    \label{fig:cdr_schemes}
\end{figure*}

In our experiments, we tested our models in two cross-domain scenarios: the dual-intra-users and the dual-inter-users. 
For the first scenario, we aim to increase recommendation quality for users inside the domain (\textit{warm-start} problem). We first filtered out unshared users to create the dataset train/test splits. Then, one item is sampled from each user history to form the test set (this method is known as \textit{leave-one-out}) as shown in Figure \ref{fig:cdr_schemes} (a). 
In the second scenario, we aim to recommend items for users "migrating" from one domain to another. It is a \textit{cold-start} problem from the point of view of the destination domain, as users do not have a history there. To emulate this scenario, we sampled the complete history of two disjoint sets of shared users from each domain, as shown in Figure \ref{fig:cdr_schemes} (b). In the cold start scenario, note that test users in one domain are considered unshared in the other since their history is sampled out. Thus, the set of users with the shared factors will be smaller than the actual intersection of users of datasets. 

As mentioned in \cite{liu2020cross}, the density of the preprocessed data was increased by filtering out users of less than 5 interactions and items with less than 10 interactions. For the warm start scenario, there are two pairs of datasets: \textbf{Sport} \& \textbf{Cloth} and \textbf{Phone} \& \textbf{Electronics}. There are also two pairs of datasets of the cold start case. These datasets are \textbf{Sport} \& \textbf{Cloth} (with different preprocessing and splitting patterns) and \textbf{Games} \& \textbf{Video}. Some statistics about the size of datasets are shown in Table \ref{tab:dataset_df}.

\begin{table}
\centering
\caption{Dataset statistics}
\label{tab:dataset_df}
\resizebox{1\columnwidth}{!}{
    \begin{tabular}{cccccccccc}
    \cline{1-9}
    \multicolumn{1}{|c|}{\multirow{2}{*}{Scenario}} &
      \multicolumn{1}{c|}{\multirow{2}{*}{Dataset}} &
      \multicolumn{4}{c|}{Train set} &
      \multicolumn{3}{c|}{Test set} &
       \\ \cline{3-9}
    \multicolumn{1}{|c|}{} &
      \multicolumn{1}{c|}{} &
      \multicolumn{1}{c|}{\#users} &
      \multicolumn{1}{c|}{\#items} &
      \multicolumn{1}{c|}{\#ratings} &
      \multicolumn{1}{c|}{\#shared} &
      \multicolumn{1}{c|}{\#users} &
      \multicolumn{1}{c|}{\#items} &
      \multicolumn{1}{c|}{\#ratings} &
       \\ \cline{1-9}
    \multicolumn{1}{|c|}{\multirow{4}{*}{\begin{tabular}[c]{@{}c@{}}intra-domain\\ (warm-start)\end{tabular}}} &
      \multicolumn{1}{c|}{Sport} &
      \multicolumn{1}{c|}{9928} &
      \multicolumn{1}{c|}{30796} &
      \multicolumn{1}{c|}{92612} &
      \multicolumn{1}{c|}{\multirow{2}{*}{9928}} &
      \multicolumn{1}{c|}{8326} &
      \multicolumn{1}{c|}{5734} &
      \multicolumn{1}{c|}{8326} &
       \\
    \multicolumn{1}{|c|}{} &
      \multicolumn{1}{c|}{Cloth} &
      \multicolumn{1}{c|}{9928} &
      \multicolumn{1}{c|}{39008} &
      \multicolumn{1}{c|}{87829} &
      \multicolumn{1}{c|}{} &
      \multicolumn{1}{c|}{7540} &
      \multicolumn{1}{c|}{5495} &
      \multicolumn{1}{c|}{7540} &
       \\ \cline{2-9}
    \multicolumn{1}{|c|}{} &
      \multicolumn{1}{c|}{Phone} &
      \multicolumn{1}{c|}{3325} &
      \multicolumn{1}{c|}{17709} &
      \multicolumn{1}{c|}{50407} &
      \multicolumn{1}{c|}{\multirow{2}{*}{3325}} &
      \multicolumn{1}{c|}{2559} &
      \multicolumn{1}{c|}{1795} &
      \multicolumn{1}{c|}{2559} &
       \\
    \multicolumn{1}{|c|}{} &
      \multicolumn{1}{c|}{Electronics} &
      \multicolumn{1}{c|}{3325} &
      \multicolumn{1}{c|}{38706} &
      \multicolumn{1}{c|}{115554} &
      \multicolumn{1}{c|}{} &
      \multicolumn{1}{c|}{2560} &
      \multicolumn{1}{c|}{2099} &
      \multicolumn{1}{c|}{2559} &
       \\ \cline{1-9}
    \multicolumn{1}{|c|}{\multirow{4}{*}{\begin{tabular}[c]{@{}c@{}}inter-domain\\ (cold-start)\end{tabular}}} &
      \multicolumn{1}{c|}{Sport} &
      \multicolumn{1}{c|}{26346} &
      \multicolumn{1}{c|}{12655} &
      \multicolumn{1}{c|}{163291} &
      \multicolumn{1}{c|}{\multirow{2}{*}{25356}} &
      \multicolumn{1}{c|}{897} &
      \multicolumn{1}{c|}{2698} &
      \multicolumn{1}{c|}{3546} &
       \\
    \multicolumn{1}{|c|}{} &
      \multicolumn{1}{c|}{Cloth} &  
      \multicolumn{1}{c|}{40839} &
      \multicolumn{1}{c|}{17943} &
      \multicolumn{1}{c|}{187880} &
      \multicolumn{1}{c|}{} &
      \multicolumn{1}{c|}{907} &
      \multicolumn{1}{c|}{2519} &
      \multicolumn{1}{c|}{3085} &
       \\ \cline{2-9}
    \multicolumn{1}{|c|}{} &
      \multicolumn{1}{c|}{Video} &
      \multicolumn{1}{c|}{19240} &
      \multicolumn{1}{c|}{8751} &
      \multicolumn{1}{c|}{156091} &
      \multicolumn{1}{c|}{\multirow{2}{*}{19014}} &
      \multicolumn{1}{c|}{216} &
      \multicolumn{1}{c|}{1157} &
      \multicolumn{1}{c|}{1458} &
       \\
    \multicolumn{1}{|c|}{} &
      \multicolumn{1}{c|}{Games} &
      \multicolumn{1}{c|}{24799} &
      \multicolumn{1}{c|}{12319} &
      \multicolumn{1}{c|}{155036} &
      \multicolumn{1}{c|}{} &
      \multicolumn{1}{c|}{212} &
      \multicolumn{1}{c|}{1138} &
      \multicolumn{1}{c|}{1304} &
       \\ \cline{1-9}
    \end{tabular}
    }
\end{table}

\subsection{Evaluation Method and Metrics}
Sampled metrics are getting increasingly popular in many research works. They provide faster computations as relevant items are ranked against a random sample of negative items, not the entire catalog. However, they may be biased when comparing different models \cite{krichene2020sampled}. However, with a sufficiently large sample size, we claim to have an unbiased estimation of model performance. In the following experiments, we sampled 999 negative items for each item in the test set. Then, we rank this item list and evaluate the top 10 recommended items with the two famous metrics: \textit{NDCG} (Normalized Discounted Cumulative Gain) and \textit{HR} (Hit Ratio).

We used a separate validation set to tune the hyperparameters. In the \href{https://drive.google.com/drive/folders/1DCYiFU6GCVj681GKYUY2d_BJFln1-8gL}{preprocessed datasets} provided by \cite{Cao2023}, there are separate files for validation sets for each domain in the warm-start scenario, but not for the cold-start. Thus, for the latter scenario, we split the provided training set into training and validation sets. We also used the \emph{leave\_one\_out} method to make this validation set, where one item is sampled out of each user history. For final testing, we used the provided test sets.

\subsection{Baselines}
\paragraph{Single-domain baselines:} We first compare with BPRMF \cite{rendle2012bpr}, NeuMF \cite{he2017neural}. These are typical baseline methods for industry because of the straightforward idea and simple architecture. 
The latest graph-based methods were included in the comparison: NGCF \cite{wang2019neural} and LightGCN \cite{he2020lightgcn}. These models aggregate the high-order neighbor information for the user/item. For single-domain methods, all domain interaction data were mixed as one domain to train them.

However, CDIMF is a cross-domain variant of ALS \cite{Rendle2021}. Thus in both scenarios, we report the results of a single ALS model trained on the joined datasets. We joined the training dataframes so the resulting dataframe has a number of users equal to the size of the union of user sets and a number of items equal to the sum of the numbers of items in both domains since the item sets are disjoint. This baseline aims to show the performance of full cross-domain sharing with no privacy restrictions.

For the warm-start specifically, we report another baseline. It is the ALS trained on each domain separately. This setting is equivalent to running the CDIMF code with sharing parameter $\rho=0$. It would help understand whether sharing knowledge across domains has a positive or negative impact on recommendation quality compared to the single domain case. In the cold start, there is no point in comparing with ALS on separate datasets since user factors on each domain would be learned independently, and transferring these factors to other domains will result in systematically wrong recommendations. 

\paragraph{Intra-domain CDR baselines:} For this scenario, we compare with MLP-based models like CoNet \cite{hu2018conet} and DDTCDR \cite{li2019ddtcdr}. These models assign an encoder for each domain and learn different dual transferring modules between source and target domains. In addition, recent GNN-based models like Bi-TGCF \cite{liu2020cross} and DisenCDR \cite{cao2022disencdr} are included in the comparison. We also compare with UniCDR \cite{Cao2023}, a model that follows a contrastive learning approach and scores good metrics.

\paragraph{Inter-domain CDR baselines:} For this category of CDR methods, the models SSCDR \cite{kang2019semi}, TMCDR \cite{zhu2021transfer}, and SA-VAE \cite{salah2021towards} mainly rely on the learning embedding and then mapping them. They train an alignment function to project the user embeddings from the source domain to the target domain. Besides, CDRIB \cite{cao2022cross} depends on the variational information bottleneck principle to learn an unbiased representation to encode the domain invariant information. We also compare with UniCDR \cite{Cao2023} as it is a model designed to work in both cold-start and warm-start settings.

\section{Results}
\label{sec:results}
We report the results of our experiments in this section. The first set of experiments aims to compare the proposed model with the baselines, and the second group of experiments explores the behavior of CDIMF and the effect of different hyperparameters on it.

\subsection{Performance Comparison}
Table \ref{tab:cdr_intra_results} reports the performance metrics of the proposed model and the compared models for warm start. CDIMF demonstrates significant improvement in most experiments for this scenario. However, these advancements are more noticeable in some experiments than in others. Although CDIMF fails to score the top HR@10 on some datasets, it scores the best NDCG@10. The latter metric is more critical and less noisy than the Hit Rate since it provides a more comprehensive evaluation of the overall quality of recommendations and how well relevant items are positioned in the list. The results of the ALS model trained on separate datasets demonstrate that sharing cross-domain knowledge does not guarantee better performance in some experiments. One example is in Table \ref{tab:cdr_intra_results}, where ALS achieved the highest metrics for the \textbf{Phone} domain in the second experiment. 

Table \ref{tab:cdr_inter_results} shows the performance metrics of the CDIMF and the compared models for the cold start scenario. CDIMF shows a significant increase in metrics in most experiments in this scenario as well. However, these improvements are less significant than in warm-start experiments. We also can notice that the ALS on the joined dataset score less than the top performing baselines in the \textbf{Game-Video} experiment, even less than the CDIMF, unlike other experiments. That may reflect that this pair of datasets in particular contains patterns that can't be learned using models based on simple matrix factorization.

 
One general observation that can be concluded from both tables is that CDIMF scores are slightly lower than the ALS model trained on the mixed dataset, e.g., incorporated data from both domains into a single dataset. \emph{These correlated scores support our claim that CDIMF is a cross-domain variant of ALS.} Unlike ALS on the joint datasets, CDIMF does not share explicit user data between domains.

The last point to mention is that basic single-domain methods trained on each domain dataset separately show competing performance to cross-domain models in some experiments. This observation raises the question of whether sharing cross-domain knowledge is always useful for all pairs of domains. This question may establish an interesting research topic about the "compatibility" of datasets and how to estimate the expected benefit of cross-domain knowledge sharing.
\begin{table*}
\caption{Performance comparison of our proposed model on dual-intra-domain (warm start) recommendations.The asterisk sign indicates that the single-domain method trained separately on local data outperformed all cross-domain methods.}
\label{tab:cdr_intra_results}
    \begin{tabular}{cc|cccc|cccc}
    \multicolumn{2}{c|}{Datasets} &
      \multicolumn{2}{c}{Sport} &
      \multicolumn{2}{c|}{Cloth} &
      \multicolumn{2}{c}{Electronic} &
      \multicolumn{2}{c}{Phone} \\ \hline
    \multicolumn{2}{c|}{Metrics@10} &
      HR &
      NDCG &
      HR &
      NDCG &
      HR &
      NDCG &
      HR &
      NDCG \\ \hline
    \multicolumn{1}{c|}{\multirow{5}{*}{\begin{tabular}[c]{@{}c@{}}Single-domain\\ methods\end{tabular}}} &
      BPRMF &
      10.43 &
      5.41 &
      11.53 &
      6.25 &
      15.71 &
      9.19 &
      16.32 &
      8.53 \\
    \multicolumn{1}{c|}{} &
      NeuMF &
      10.74 &
      5.46 &
      11.18 &
      6.02 &
      16.17 &
      9.24 &
      15.84 &
      8.02 \\
    \multicolumn{1}{c|}{} &
      LightGCN &
      13.19 &
      6.94 &
      13.58 &
      7.29 &
      19.17 &
      10.28 &
      23.25 &
      12.72 \\
    \multicolumn{1}{c|}{} &
      ALS (separate)&
      16.67 &
      10.43 &
      13.9 &
      8.86 &
      22.11 &
      14.01 &
      29.54* &
      18.74* \\ 
    \multicolumn{1}{c|}{} &
      ALS (joined)&
      \textit{25.69} &
      \textit{18.44} &
      \textit{22.15 }&
      \textit{16.4} &
      \textit{21.37} &
      \textit{14.91} &
      \textit{32.51 }&
      \textit{22.64} \\ \hline
    \multicolumn{1}{c|}{\multirow{5}{*}{\begin{tabular}[c]{@{}c@{}}Cross-domain\\ methods\end{tabular}}} &
      CoNet &
      12.09 &
      6.41 &
      12.40 &
      6.62 &
      17.22 &
      9.86 &
      17.66 &
      9.30 \\
    \multicolumn{1}{l|}{} &
      DDTCDR &
      11.86 &
      6.37 &
      12.54 &
      7.13 &
      18.47 &
      11.08 &
      17.23 &
      8.58 \\
    \multicolumn{1}{l|}{} &
      Bi-TGCF &
      14.83 &
      7.95 &
      14.68 &
      7.93 &
      22.14 &
      12.20 &
      25.71 &
      13.93 \\
    \multicolumn{1}{l|}{} &
      DisenCDR &
      17.55 &
      9.46 &
      16.31 &
      9.03 &
      \textbf{24.57} &
      \underline{14.51} &
      \textbf{28.76} &
      \underline{16.13} \\
    \multicolumn{1}{l|}{} &
      UniCDR &
      \underline{18.37} &
      \underline{10.98} &
      \underline{17.85} &
      \underline{11.20} &
      \underline{22.92} &
      13.83 &
      24.72 &
      13.77 \\ \hline
    \multicolumn{1}{c|}{\multirow{1}{*}{Our method}} &
      CDIMF &
      \textbf{23.85} &
      \textbf{17.12} &
      \textbf{20.74} &
      \textbf{15.11} &
      22.66 &
      \textbf{14.65} &
      \underline{28.57} &
      \textbf{18.3}
    \end{tabular}
    
\end{table*}
\begin{table*}
\caption{Performance comparison of our proposed model on dual-inter-domain (cold start) recommendations}
\label{tab:cdr_inter_results}
    \begin{tabular}{cc|cccc|cccc}
    \multicolumn{2}{c|}{Datasets} &
      \multicolumn{2}{c}{Sport} &
      \multicolumn{2}{c|}{Cloth} &
      \multicolumn{2}{c}{Game} &
      \multicolumn{2}{c}{Video} \\ \hline
    \multicolumn{2}{c|}{Metrics@10} & HR          & NDCG       & HR          & NDCG       & HR   & NDCG & HR    & NDCG \\ \hline
    \multicolumn{1}{c|}{\multirow{3}{*}{\begin{tabular}[c]{@{}c@{}}Single-domain\\ methods\end{tabular}}} &
      BPRMF &
      5.75 &
      3.16 &
      6.75 &
      3.26 &
      3.77 &
      1.89 &
      4.46 &
      2.36 \\
    \multicolumn{1}{c|}{}  & 
    NGCF & 
    7.22 & 
    3.63 & 
    7.07 & 
    3.48 & 
    5.14 & 
    2.73 & 
    7.41 & 
    3.87 \\ 
    \multicolumn{1}{c|}{} &
      ALS (joined)&
      \textit{15.01} &
      \textit{10.61} &
      \textit{13.68}&
      \textit{9.36} &
      \textit{7.44} &
      \textit{3.83} &
      \textit{8.16}&
      \textit{4.64} \\ \hline
    
    \multicolumn{1}{c|}{\multirow{5}{*}{\begin{tabular}[c]{@{}c@{}}Cross-domain\\ methods\end{tabular}}} &
      SSCDR(CML) &
      7.27 &
      3.75 &
      6.12 &
      3.06 &
      3.48 &
      1.59 &
      5.51 &
      2.61 \\
    \multicolumn{1}{c|}{}  & TMCDR  & 7.18        & 3.84       & 8.11        & 5.05       & 5.36 & 2.58 & 8.85  & 4.41 \\
    \multicolumn{1}{c|}{}  & SA-VAE & 7.51        & 3.72       & 7.21        & 4.59       & 5.84 & 2.78 & 7.46  & 3.71 \\
    \multicolumn{1}{c|}{}  & CDRIB  & \underline{12.04} & 6.22       & 12.19 & 6.81       & 8.51 & \underline{4.58} & \textbf{13.17} & \textbf{6.49} \\
    \multicolumn{1}{c|}{}  & UniCDR & 11.20       & \underline{7.04} & \underline{12.48}       & \underline{7.52} & 8.78 & \textbf{4.63} & \underline{10.74} & \underline{5.89} \\ \hline
    \multicolumn{1}{c|}{Our method} &
      CDIMF &
      \textbf{14.52} &
      \textbf{9.51} &
      \textbf{13.35} &
      \textbf{9.45} &
      \textbf{9.19} &
      4.53 &
      8.50 &
      4.36
    \end{tabular}
\end{table*}

\subsection{Hyperparameters Discussion}
Apart from the hyperparameters of the ALS solver described in Section \ref{subsec:als_adapt}, CDIMF has three characteristic hyperparameters: the sharing (penalty) parameter, the proximal operator, and the aggregation period. We conducted several experiments to understand the effect of each hyperparameter on the behavior of CDIMF.

\subsubsection{The effect of sharing parameter $\rho$}
\label{subsec:rho_effect}
\begin{figure}
    \centering
    \includegraphics[width=\columnwidth]{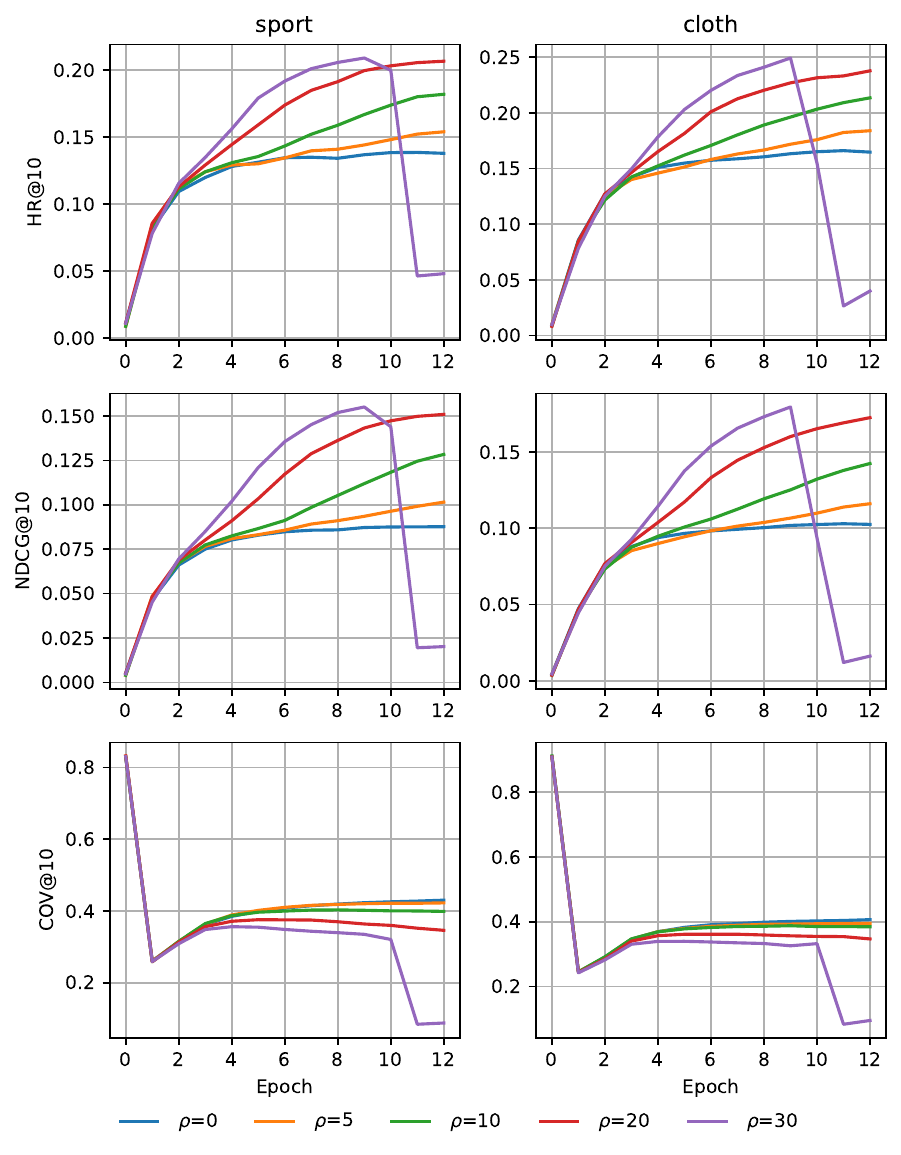}
    \caption{HR@10, NDCG@10, and coverage changes in the warm start problem for several values of sharing parameter $\rho$. For $\rho=0$, a single domain ALS is running on each side.} 
    \Description[Metric changes for values of sharing parameter]{Metric changes for values of sharing parameter}
    \label{fig:warm_vs_rho}   
\end{figure}
In Figure \ref{fig:warm_vs_rho}, we can see that the sharing factor $\rho$ plays a role in the learning rate of CDIMF as it controls the learning speed. For $\rho=0$ the iteration in \eqref{eq:admm_update_local} would be:
\begin{equation}
    X_i^{*}, Y_i^{*} = \argmin_{X_i,Y_i} \|C_i \circ (P_i-X_i Y_i^T)\|_F^2 + \|\Lambda_{Xi} X_i\|_F^2+ \|\Lambda_{Yi} Y_i\|_F^2, \nonumber
\end{equation}
which is simply a single-domain ALS. The model learns cross-domain knowledge more rapidly as the sharing parameter increases. The model starts to diverge for higher values of $\rho$, leading to the learning of divergent item factors. This results in a few items with the largest norms, which score larger dot products with any user factor. Therefore, these items are usually recommended. This explains why the coverage drops dramatically to a near-zero value when divergence occurs.


\subsubsection{The effect of proximal operator}
\begin{figure}
    \centering
    \includegraphics[width=\columnwidth]{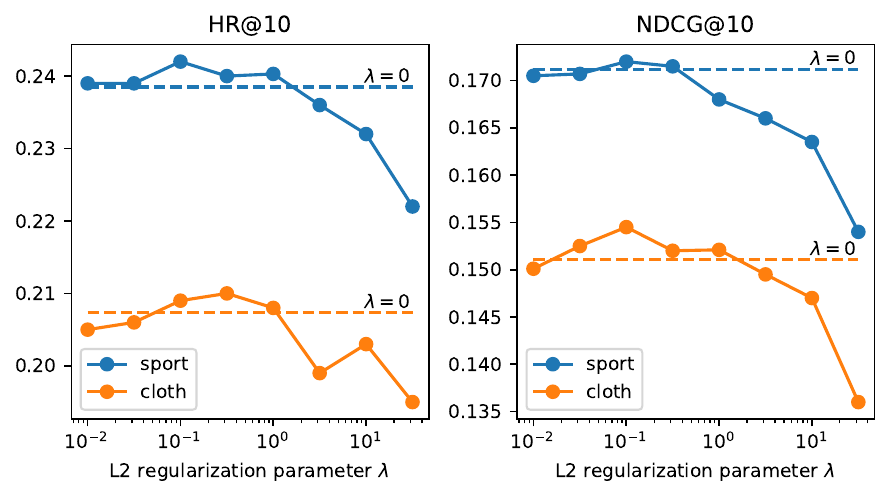}
    \caption{The effect of \textit{L2} regularization on performance of CDIMF. Scenario: warm start, dataset: \textbf{Cloth\&Sport}}
    \Description[Effect of \textit{L2} regularization]{Effect of \textit{L2} regularization}
    \label{fig:warm_vs_reg}
\end{figure}
In this paper, we considered two simple options for the proximal operator in \eqref{eq:admm_update_global}; the identity function and the \textit{L2} regularization. Figure \ref{fig:warm_vs_reg} demonstrates the effect of \textit{L2} regularization on the CDIMF models. Although choosing the optimal value for the regularization parameter $\lambda$ can enhance the performance slightly over the case of no regularizer, reported changes are relatively small. Finding the best value of $\lambda$ increases the time of hyperparameter tuning without significant improvement. Thus, all numbers in Tables \ref{tab:cdr_intra_results} and \ref{tab:cdr_inter_results} are for the CDIMF with identity operator.  
\subsubsection{The effect of aggregation period}
\begin{figure}
    \centering
    \includegraphics[width=0.95\columnwidth]{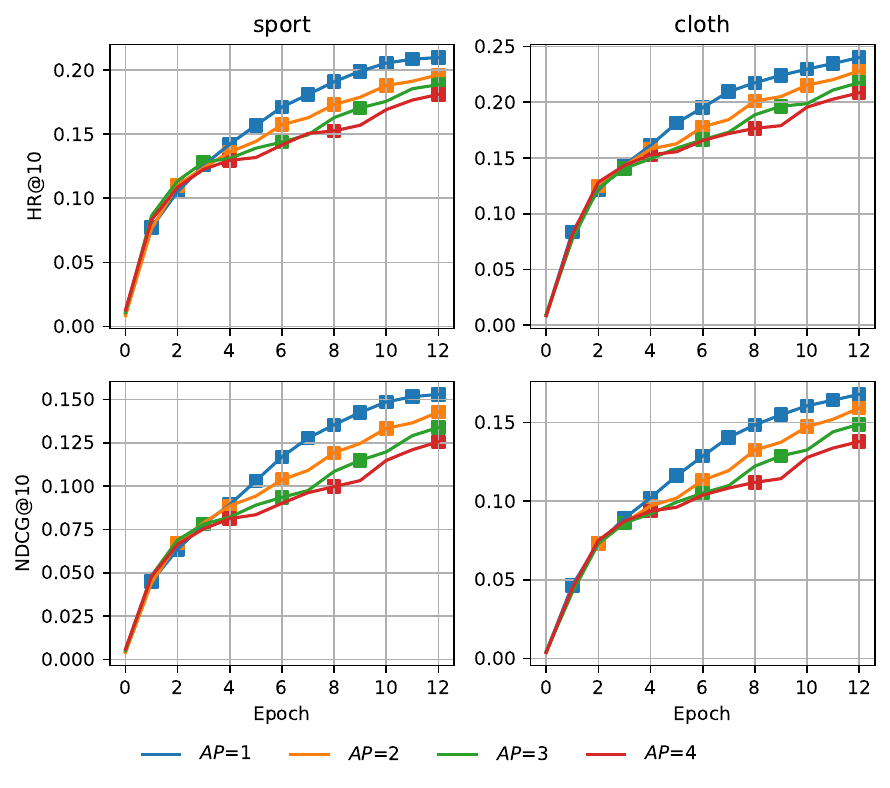}
    \caption{HR@10 and NDCG@10 changes in the warm start problem for several values of aggregation period (AP), squares refer to epochs of shared variable update}
    \label{fig:warm_vs_agg}
    \Description[Metric changes by epochs]{Metric changes by epochs}
\end{figure}

The typical ADMM iterations assume having the optimal solution locally \eqref{eq:admm_update_local} before updating the common variable \eqref{eq:admm_update_global}, experiments show that sharing a sub-optimal solution (resulting from a single ALS epoch) frequently is better. However, the communication load would increase in this situation. Figure \ref{fig:warm_vs_agg} shows that the more frequent aggregations are, the faster performance growth is. This finding may allow asynchronous updates of the global variable. Supposing that the global variable $Z$ was not updated in some epochs due to communication issues, a local ALS iteration could be run using the old version of $Z$ and still achieve decent, yet lower, improvement. 

\subsection{Privacy Considerations}

The variable shared by the node in each domain is $X_i+U_i$, which is a mixture of the local factors of the users and the dual variable related to the equality constraint $X_i - Z=0$. We hypothesize that the explicit user history in a particular domain cannot be extracted unless the item factors of that domain are exposed. Thus, communication could be held over an insecure channel as long as the domains' servers are well protected. However, further investigation of the possible reconstruction attacks is needed in this aspect. 

\section{Conclusion}
\label{sec:conclusion}
In this paper, we introduced CDIMF, a model that upgrades the conventional ALS method to the cross-domain context. Our experimental results demonstrate that CDIMF outperforms other state-of-the-art CDR methods in most cases, often surpassing even more elaborate neural network-based approaches. Consequently, among the key advantages of our CDIMF approach are competitive performance and superior computational efficiency inherited from ALS, which makes it plausible for practical use cases.


CDIMF model has been tested in the dual domain case for both warm-start and cold-start problems. Moreover, the extension to the multi-domain case can be achieved with minor adjustments. The task here is to deal with cases of users who have a history in more than one, but not all, domains. Further experiments should be conducted to evaluate possible solutions for this non-homogenous setting. The CDIMF model also has the advantage of notation symmetricity, and it can extend to item-overlapping scenarios by swapping the notations.


\bibliographystyle{ACM-Reference-Format}
\bibliography{content/90_bib}

\end{document}